\pdfoutput=1
\documentclass[12pt]{article}

\usepackage{graphicx}
\usepackage{xcolor}
\usepackage{fancyhdr}
\usepackage{ae,aecompl}
\usepackage{epsfig}
\usepackage{mathtools}
\usepackage{ mathrsfs }
\usepackage{ upgreek }
\usepackage{verbatim}
\usepackage{IEEEtrantools}
\usepackage{caption}
\usepackage{amsfonts}
\usepackage{amssymb}
\usepackage{graphicx}
\usepackage{lscape}
\usepackage{amsmath}
\usepackage{caption}
\usepackage{subcaption}
\usepackage{textcomp}
\usepackage{txfonts}
\usepackage{pstricks}
\usepackage{ dsfont }
\usepackage{pstricks}
\usepackage{multirow}
\usepackage{booktabs}
\usepackage{graphics}
\usepackage{array}
\usepackage{verbatim}
\usepackage{float}
\usepackage{titlesec}
\usepackage{multirow}
\usepackage[english]{babel}
\usepackage{dcolumn}
\usepackage{tikz}
\usepackage[utf8]{inputenc}

\begin{document}


\title{Neutrino mass hierarchy and $\delta^{CP}$ investigation within the biprobability ($P-P^{T}$) plane}
\author{Mandip Singh$^*$ \\
\\
{ \it Department of Physics, Centre of Advanced Study, P. U., Chandigarh, India.}\\
}

\maketitle
\begin{abstract}
This article illustrates the possibility of investigating mass hierarchy and CP-violating phase $\delta^{CP}$,
in the context of CP trajectory diagrams in the bi-probability plane.
Separation between normal mass hierarchy (NH) and inverted mass hierarchy (IH) 
CP trajectory ellipses in the $P-P^{T}$ plane seems to be 
very promising in order to investigate mass hierarchy.
Illustration of separation between two hierarchy ellipses in the E-L plane is very helping to cover all
the desired base lines and beam energies and also to analyze benefits and drawbacks at single place.
If we know the mass hierarchy, 
then from the large sizes of CP trajectory ellipse which is possible 
at appropriately long base line (L) and at specific value of beam energy (E), 
it becomes possible to investigate at-least narrow
ranges of CP/T-violating phase $\delta^{CP}$. The Possibility of more 
than one set of ($\theta_{13}, ~\delta^{CP}$) 
parameters to correspond to any chosen coordinate in
$P-P^{T}$ plane, known as parameter degeneracy, 
may hinder exact determination of mass hierarchy as well as $\delta^{CP}$ value. 
To circumvent this degeneracy in the ($\theta_{13}, ~\delta^{CP}$) parameter space, 
in case of opposite sign solutions corresponding to 
NH and IH case points toward the need of sufficiently long base lines, 
so as to separate opposite hierarchy 
ellipses to observable separation, and in case of same sign solutions 
corresponding to either NH or IH, 
we need to choose an experimental configuration 
with $L \simeq 2,535$ Km, $E \simeq 5$ GeV for n=1 scenario.
\end{abstract}

\newpage
\maketitle

\section{Introduction}
Though during past few decades neutrino oscillation experiments \cite{t2k}-\cite{opera}
have made great success in revealing the structure of lepton flavor mixing matrix, but still there remain several 
unanswered questions. Investigation of the fact, whether third neutrino is heavier (lighter) than the two neutrinos, whose 
mass square difference is responsible for atmospheric neutrino oscillations i.e. $\Delta m_{31}^{2} \simeq \Delta m_{32}^{2} \ge 0$ 
($\Delta m_{31}^{2} \simeq \Delta m_{32}^{2} \le 0$), known as normal (inverted) mass hierarchy, is one of the most intriguing unanswered  
questions till now. Investigation of mass hierarchy would be very helpful to disclose the theoretical phenomenology describing the 
neutrino mass. An another tantalizing question that present and future oscillation experiments have to uncover is the 
determination of leptonic CP-violation strength, which depends on CP/T-violation phase $\delta^{CP}$ or $\delta^{T}$. 
This will not only shed light on the theoretical structure behind the complete parallelism between quarks and leptons,
but also may be very useful in understanding matter anti-matter asymmetry arising due to baryon number violation in the universe \cite{basym}.

An expected small value of $\theta_{13}$ mixing angle is the major problem, as it obscure the sensitivity of 
both CP-violating phase $\delta^{CP}$ and mass hierarchy effects \cite{th13n}, as is also evident from Eqn's.~(\ref{parti}) and (\ref{coeff}).
Recent confirmation of moderately large value of reactor mixing angle ($\theta_{13} \sim 9^{o}$)
\cite{th13a}, \cite{th13b} has opened the door of the possibility for precise determination of leptonic T/CP phase $\delta^{CP}$. 

Due to the fact that $\Delta P_{e \mu}^{T} = \Delta P_{\mu \uptau}^{T}
= \Delta P_{\uptau~ e}^{T} = -\Delta P_{\mu~ e}^{T} = -\Delta P_{\uptau \mu}^{T} = -\Delta P_{e \uptau}^{T}$ \cite{tvilekh}, \cite{tvilpet}, 
which is true for the constant matter density approximation,
sizes and separation between ellipses in $P-P^{T}$ plane will be same for all appearance channels in three flavor oscillations, hence
it is enough to study any one suitable channel. Relatively long lived charged particles $\pi^{\pm}$ mesons can 
be easily stored in accelerators and can be accelerated to the desired energies, which make them the most easily available source of 
$\nu_{\mu}$ and $\nu_{e}$ beams. Due to this reason present and most of future oscillation experiments are 
operating with $\nu_{e} \leftrightarrow \nu_{\mu}$ channel, to which have been preferred to discuss in this article. In order to 
complete our investigation toward the investigation of optimized experimental configurations (i.e. base line L and beam energy E),
we will consider experiments viz. Brookhaven-Cornell (L=350 Km ), CERN-Gran Sasso (L=730 Km), JHF-Seoul (L=1200 Km ),
JHF-Beijing (L=2100 Km) and Fermilab-SLAC (L=2900 Km). We will restrict to 
base lines $\le 3000$ Km, so as to fulfill the constant matter density approximation, as below this 
base line value we remain confined to Earth's crust layer region only.
Experiment with smallest base line L=350 Km assume the 
approximation of vacuum oscillations, because of very small matter effects. 
At long base lines matter contamination to the 
oscillation effects becomes significantly large, which fake the signal arising due to true/intrinsic T-violation phase in such experiments. 
Though contamination of matter effects hinders the precise direct measurement of CP-violation phase, 
but it separates the ellipses corresponding to 
NH and IH far apart, which makes the hierarchy 
investigation possible. This effect has
been illustrated in Figure~\ref{fig:elpsep}.
Matter contamination at long base lines also increases
the size of ellipse at specific beam energy, which in turn make the possible 
determination of narrow range for $\delta_{CP}$ phase, this has been illustrated in 
Figure~\ref{fig:majmin}. In this article where ever we will study any oscillation 
effect at specific value of beam energy, we will consider it as the average 
beam energy.

There exists a four fold degeneracy in the $(\theta_{13}, ~\delta^{CP})$ parameter space 
for any given coordinate in $P-P^{T}$ plane \cite{pdegn1}, \cite{pdegn2}, \cite{min3}.
The values of 
more than one
$\theta_{13}$ related solutions 
especially in case of small base lines,
may lie in the 2$\sigma$ or 3$\sigma$ range 
of this parameter, which unable the exact determination
of phase $\delta^{CP}$. Last section discusses in detail of the sources,
and how to circumvent this degeneracy within the currently available $3\sigma$ ranges of mixing parameters \cite{parabf}. 
\section{T-violation phenomenology}
\label{sec:phenology}
The neutrino oscillation pattern in vacuum can be modified significantly during the passage
of neutrinos through matter because of the effect of coherent forward scattering of
the neutrinos on the nucleons, as has been pointed out 
by  Wolfenstein \cite{wolf} and by Mikheyev and Smirnov \cite{simrn}.
We can find neutrino transition probability in matter by starting from neutrino 
oscillation in vacuum with consideration of the evolution 
of the neutrino state vector described by "Schr$\ddot{o}$dinger equation", 
for literature may see \cite{lit1}-\cite{lit7}.
Throughout the whole literature, results of neutrino oscillation experiments
have been usually analyzed 
under the simplest assumption of oscillations between two neutrino types. 
We can formulate the transition of neutrino
flavor $\nu_{e}$ to the flavor $\nu_{\mu}$ \cite{seri}, 
also known as Golden channel to the form as 
\begin{eqnarray}
 P_{e \mu}^{\pm} &=& \eta 1^{\pm} + ~ \eta 2^{\pm}~cos~\delta^{CP} ~ \pm ~ \eta 3^{\pm}~ sin~\delta^{CP}   
 \label{parti}
\end{eqnarray}
where upper sign corresponds to neutrinos and lower sign to anti-neutrinos, s.t.
\begin{subequations}
\begin{eqnarray}
 P_{e \mu}^{+} &\equiv& P(\nu_{e} \rightarrow \nu_{\mu})    \nonumber \\  
   &=& \eta 1^{+} + ~ \eta 2^{+}~cos~\delta^{CP} ~ + ~ \eta 3^{+}~ sin~\delta^{CP} 
 \label{parti1}   \\  
 P_{e \mu}^{-} &\equiv& P(\overline{\nu}_{e} \rightarrow \overline{\nu}_{\mu})  \nonumber  \\ 
   &=& \eta 1^{-} + ~ \eta 2^{-}~cos~\delta_{CP} 
     ~ - ~ \eta 3^{-}~ sin~\delta^{CP}   
 \label{anti1} 
\end{eqnarray}
\end{subequations}
In the case of symmetric density profile, we have $P_{\mu e}= P_{e \mu}(\delta^{CP} \rightarrow - \delta^{CP})$.
Thus from Eqn.~(\ref{parti}), we can write 
\begin{eqnarray}
 P_{\mu e}^{\pm} &=& \eta 1^{\pm} + ~ \eta 2^{\pm}~cos~\delta^{CP} ~ \mp ~ \eta 3^{\pm}~ sin~\delta^{CP}   
 \label{partirev}
\end{eqnarray}
If we compare Eqn's.~(\ref{parti}) and (\ref{partirev}), we can define following relations
\begin{eqnarray}
  P_{e \mu}^{+} =  P_{\mu e}^{-} ~~~ and ~~~ P_{e \mu}^{-} =  P_{\mu e}^{+}
  \label{tvileq}
\end{eqnarray}
where the coefficients has been defined as
\begin{eqnarray}
 \eta 1^{\pm} &=&  \alpha^{2} ~sin^{2} {2\theta_{12}}~ c^{2}_{23}~ \frac{sin^{2}{[A \Delta \frac{L}{2}]}}{A^{2}} \nonumber  \\ 
 &&  + 4~ s^{2}_{13}~ s^{2}_{23} ~\frac{sin^{2}{[(A \mp 1)\Delta \frac{L}{2}]}}{(A \mp 1)^{2}}  \nonumber   \\  
 \eta 2^{\pm} &=&  2 ~\alpha~ s_{13} ~sin {~2 \theta_{12}} ~sin {~ 2\theta_{23}} \nonumber \\ 
 &&  \times \frac{sin~{[A \Delta \frac{L}{2}]}}{A} 
 ~ \frac{sin{[(A \mp 1) \Delta} \frac{L}{2}]}{(A \mp 1)} ~ cos \left[ \Delta \frac{L}{2} \right]    \nonumber   \\ 
 \eta 3^{\pm} &=& 2 ~\alpha~ s_{13} ~sin {~2 \theta_{12}} ~sin {~ 2\theta_{23}}  \nonumber \\
  && \times \frac{sin~{[A \Delta \frac{L}{2}]}}{A} ~ \frac{sin{[(A \mp 1) \Delta} \frac{L}{2}]}{(A \mp 1)} 
    ~ sin \left[ \Delta \frac{L}{2} \right]
      \label{coeff}
\end{eqnarray}  \\
with $A \equiv 2~E~V/\Delta m_{31}^{2}$, where $V = \sqrt{2} ~G_{F} ~N_{e}$; with $N_{e}$ is the number density of electrons in the medium; 
$G_{F}$ = Fermi weak coupling constant = $11.6639 \times 10^{-24} ~eV^{-2}$, $\Delta \equiv \Delta m_{31}^{2}/(2~E) \simeq 
\Delta m_{32}^{2}/(2~ E)$, L is base line length and E the beam energy and mass hierarchy parameter  $\alpha= \Delta m_{21}^{2}/ 
\Delta m_{32}^{2}$.

Hence $\eta i ~ \equiv ~ \eta i ~ (E, V, L, \Delta m_{12}^{2}, \Delta m_{13}^{2}, \theta_{12}, \theta_{13}, \theta_{23})$ with {\it i}=1, 2, 3.

Coefficients $\eta 2^{\pm}$ and $\eta 3^{\pm}$ are proportional to major-minor axis \cite{cptnu} of ellipse in $P-P^{T}$ plane with $P=P_{e \mu}$ 
and $P^{T}=P_{\mu e}$.

\begin{figure*} 
\begin{center}
\caption{An illustration to find separation between nearest edges of two hierarchy ellipses for L=730 Km and $\rho = 3.5 ~gm/cm^3$.
All the rest parameters have been chosen as the best fit values shown in table
\ref{tab:parabf}.}
\includegraphics [width=0.85\textwidth]{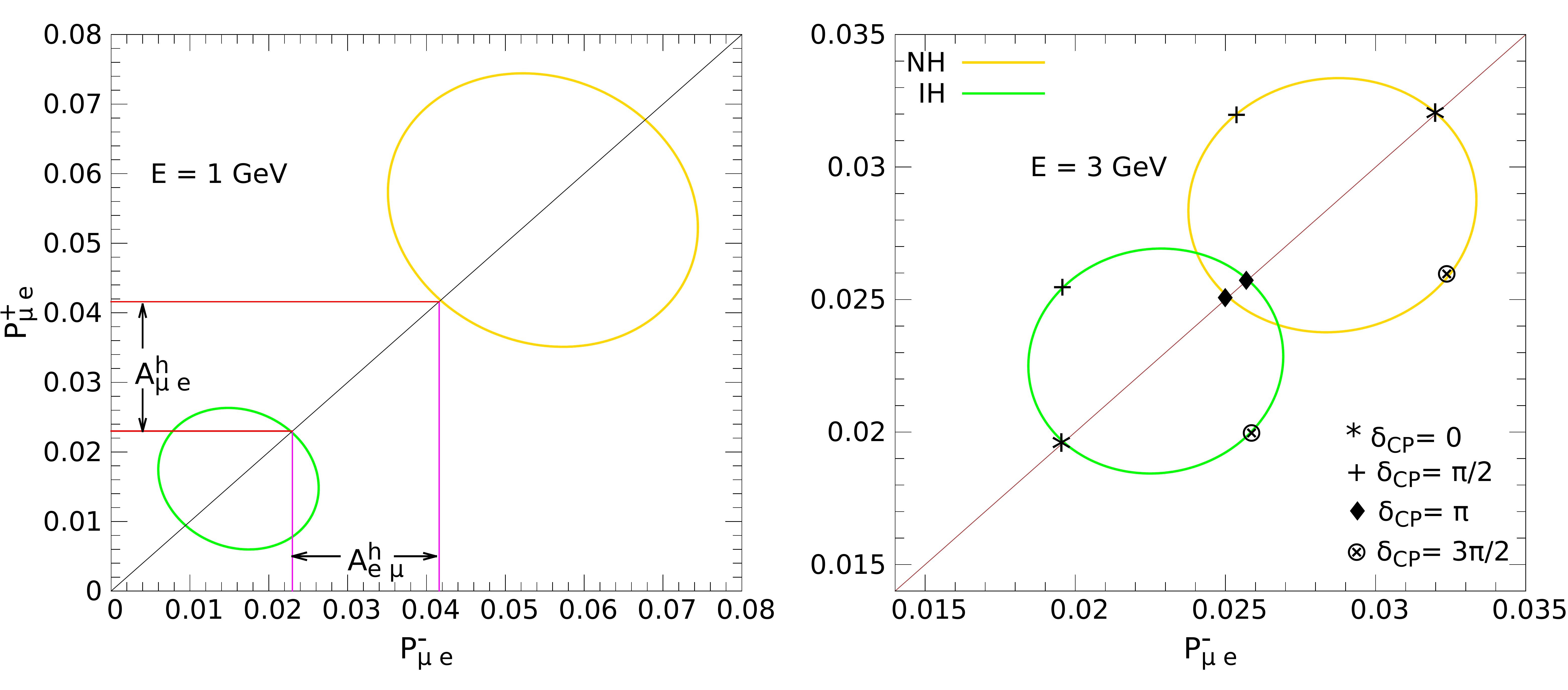}
    \label{fig:elpspr}
    \end{center}
\end{figure*}

\begin{figure*} 
\begin{center}
\caption{This figure illustrates values of nearest edges of opposite hierarchy ellipses laying along the diagonal of biprobability 
    plane (as shown in figure \ref{fig:ellips1}) and separation $A_{e \mu}^h$ (using Eqn. \ref{nprhr}) 
    between the nearest edges of these two hierarchy ellipses for experiments viz.
    Brookhaven -- Cornell (L=350 Km), CERN -- Gran Sasso ( L=730 Km), JHF -- Seoul (L=1200 Km), 
    JHF -- Beijing (L=2100 Km) and  Fermilab -- SLAC (L=2900 Km). In the constant matter density 
    approximation we have chosen $\rho=3.5 ~gm/cm^3$. Mixing parameters have been taken at their
    best fit values from table \ref{tab:parabf}.}
\includegraphics [width=.7\textwidth]{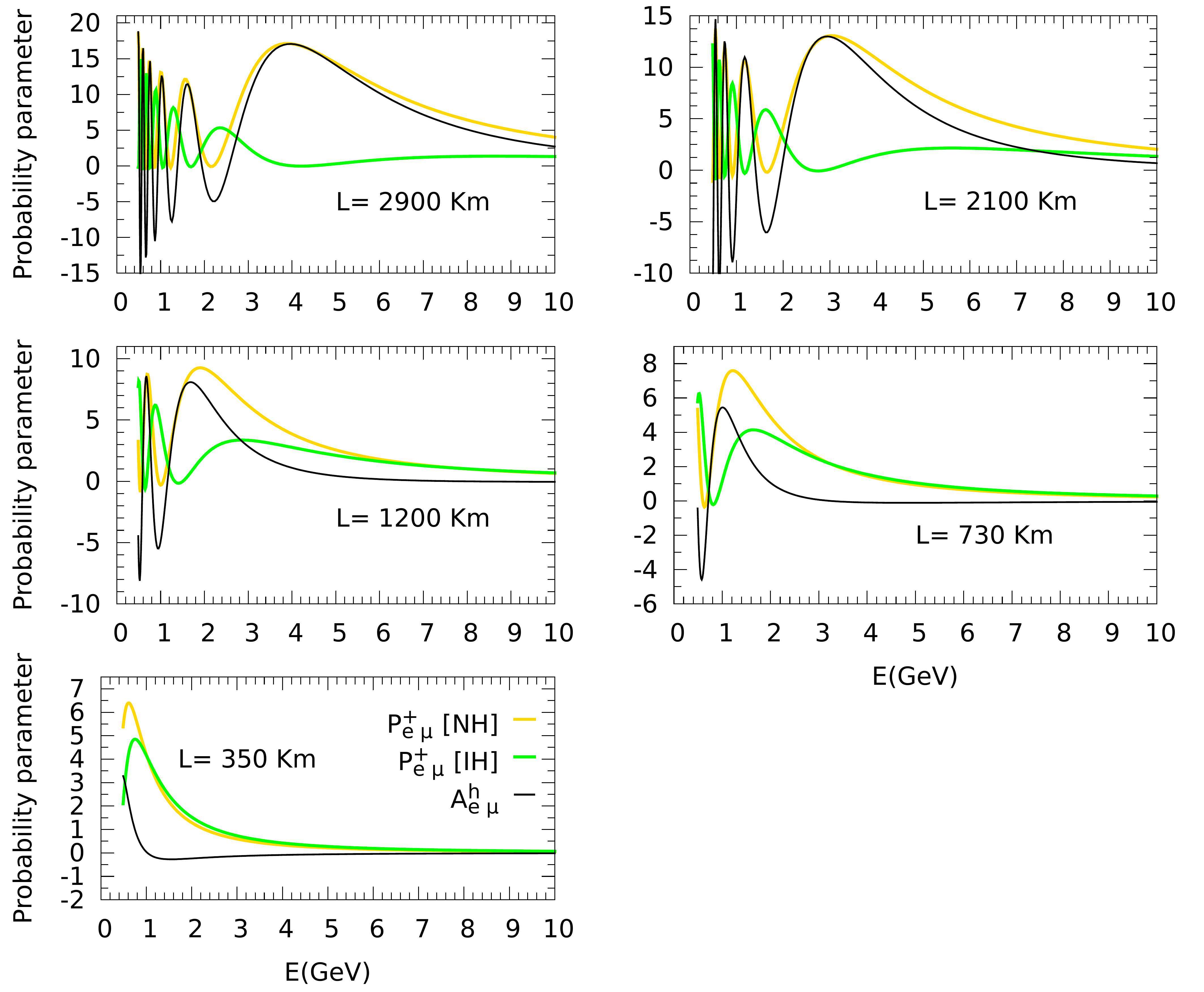}
\label{fig:ellsep2d}
\end{center}
\end{figure*}

\begin{figure*} 
\begin{center}
\caption{This Figure illustrates the parameter $A_{e \upmu}$ defined in Eqn.~(\ref{nprhr}) in the 
    E--L plane. The mixing parameters used for this figure have been tabulated in table \ref{tab:parabf}.}
\includegraphics [width=.75\textwidth]{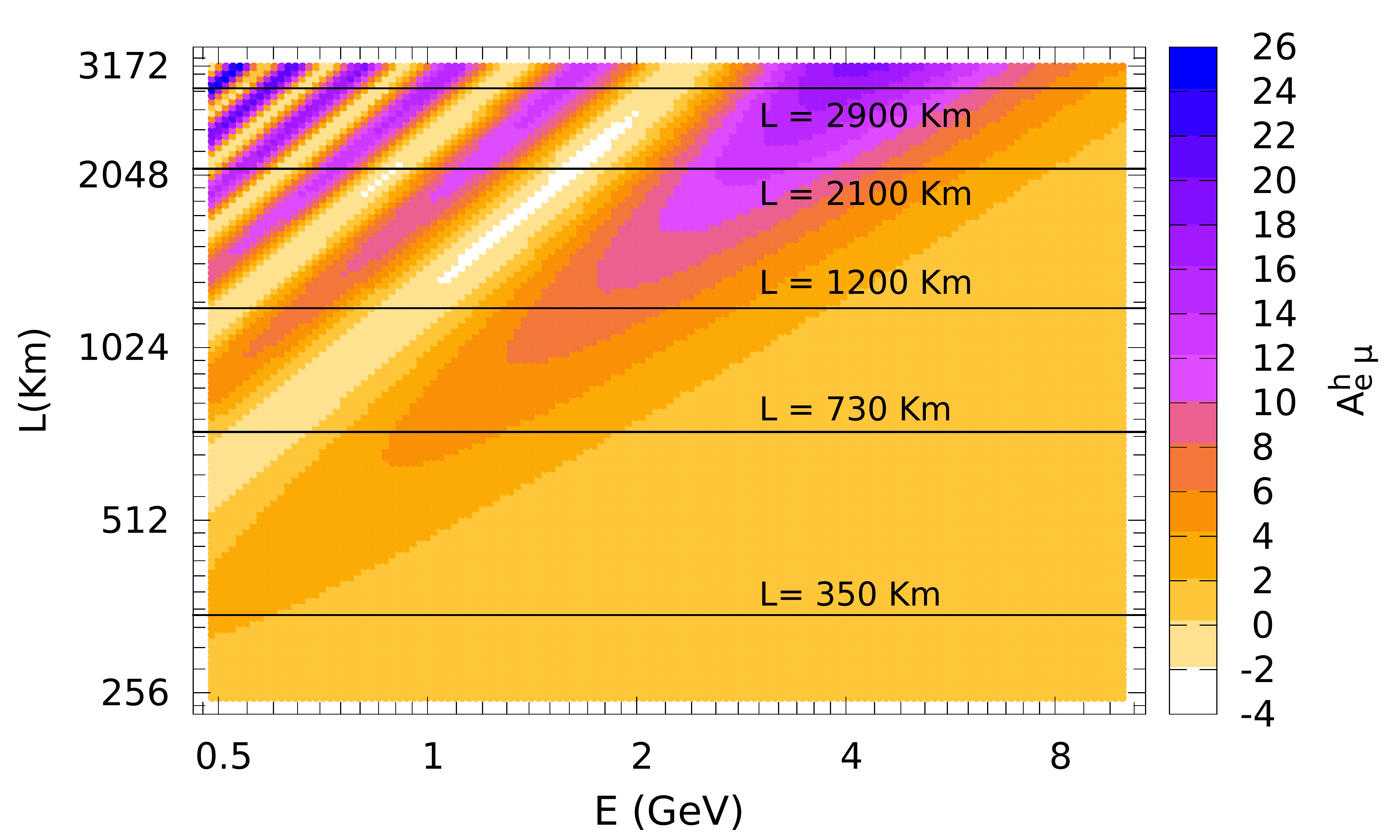}
    \label{fig:elpsep}
    \end{center}
\end{figure*}

We can define a new parameter $A^{h}_{e \mu}$ at CP-violation phase $\delta_{CP}=180^{0}$, which enables us to find the 
separation between nearby edges of NH and IH ellipses/circles centering along diagonal of $P-P^T$ 
plane as illustrated in figure \ref{fig:elpspr}. At certain value of beam energy 
two hierarchy ellipses for given base line may overlap as is evident from RHS figure.
If we bring two hierarchy ellipses close to each other these always touches firstly at
$\delta_{CP}=180^o$ point, hence it is the point 
of nearest edges. It is also evident from LHS figure that both the horizontal nearest edge 
separation (i.e. $A_{e \mu}^h$) and vertical nearest edge separation (i.e. $A_{\mu e}^h$) 
are equal i.e. $A_{e \mu}^h = A_{\mu e}^h$. We can calculate analytic expression for 
$A_{e \mu}^h$ by making use of Eqn's. (\ref{parti1}) and (\ref{coeff}) to the following form, 
\begin{eqnarray}
 A^{h}_{e \mu} & \equiv & P_{e \mu}^{+}[NH] - P_{e \mu}^{+}[IH] ~ ; ~~~~~~~~~ at ~ \delta^{CP} = 180^{o} \nonumber  \\
  &=& 4~ s^{2}_{13}~ s^{2}_{23} \left[ \frac{sin^{2}{[(A-1)\Delta \frac{L}{2}]}}{(A-1)^{2}}
   - \frac{sin^{2}{[(A+1)\Delta \frac{L}{2}]}}{(A+1)^{2}} \right]   \nonumber \\ 
     & & - ~2 ~\alpha~ s_{13} ~sin {~2 \theta_{12}} ~sin {~ 2 \theta_{23}}~ cos{\left[ \Delta \frac{L}{2} \right]}~ 
      \frac{sin~{[A \Delta \frac{L}{2}]}}{A}   \nonumber \\
      && \times  \left[ \frac{sin{[(A-1) \Delta \frac{L}{2}]}}{(A-1)} 
       - \frac{sin{[(A+1) \Delta \frac{L}{2}]}}{(A+1)} \right]
 \label{nprhr}
\end{eqnarray}
As soon as $A_{e \mu}^{h}$ or $A_{\mu e}^{h}$ attains -ve  values the two ellipses corresponding to NH and IH 
will start overlapping. This parameter has been illustrated in Figure~\ref{fig:ellsep2d} as function of beam energy E
and in Figure~\ref{fig:elpsep} in the E--L plane, known as oscillogram. 


It is evident from figure \ref{fig:ellsep2d}, the value of parameter $A_{e \mu}^h$ increases with increasing the base line length.
The one more interesting thing is that with increasing base line length position of first oscillation maximum shifts to 
higher beam energies. Shifting to higher values of beam energies benefit with increased detection cross section and more broad
maxima (or FWHM) is beneficial for better energy resolutions. It is advantageous to choose average beam energy of incident 
neutrino beam around this oscillation maximum as it has large value for parameter $A_{e \mu}^h$ along with
sufficiently large value of FWHM, which enables us to accurately measure the probability parameter.

\setlength{\arrayrulewidth}{0.31 mm}  
\setlength{\tabcolsep}{1pt}
\renewcommand{\arraystretch}{1.1}
\begin{table}[h]
\centering
\begin{tabular}{p{3.5cm} p{1.5cm} p{1.5cm}} \hline
 Parameter &  NH &  IH   \\ 
\hline
$\theta_{12}$ & $34.6^{o}$ & $34.6^{o}$   \\
$\theta_{23}$ & $48.9^{o}$ & $49.2^{o}$    \\
$\theta_{13}$ & $8.8^{o}$ & $8.9^{o}$   \\
$\Delta m_{21}^2 [10^{-5}] ~(eV^2)$  & 7.6  & 7.6   \\  
$\Delta m_{31}^2 [ 10^{-3}] ~(eV^2)$ & 2.48  & 2.38  \\ 
\hline
\end{tabular}
\captionsetup{justification=raggedright,
singlelinecheck=false
}
\caption{This table tabulates the best fit values of mixing parameters adopted from 
\cite{parabf}.}
\label{tab:parabf}
\end{table}

\begin{figure*} 
\begin{center}
\includegraphics [width=1.0\textwidth] {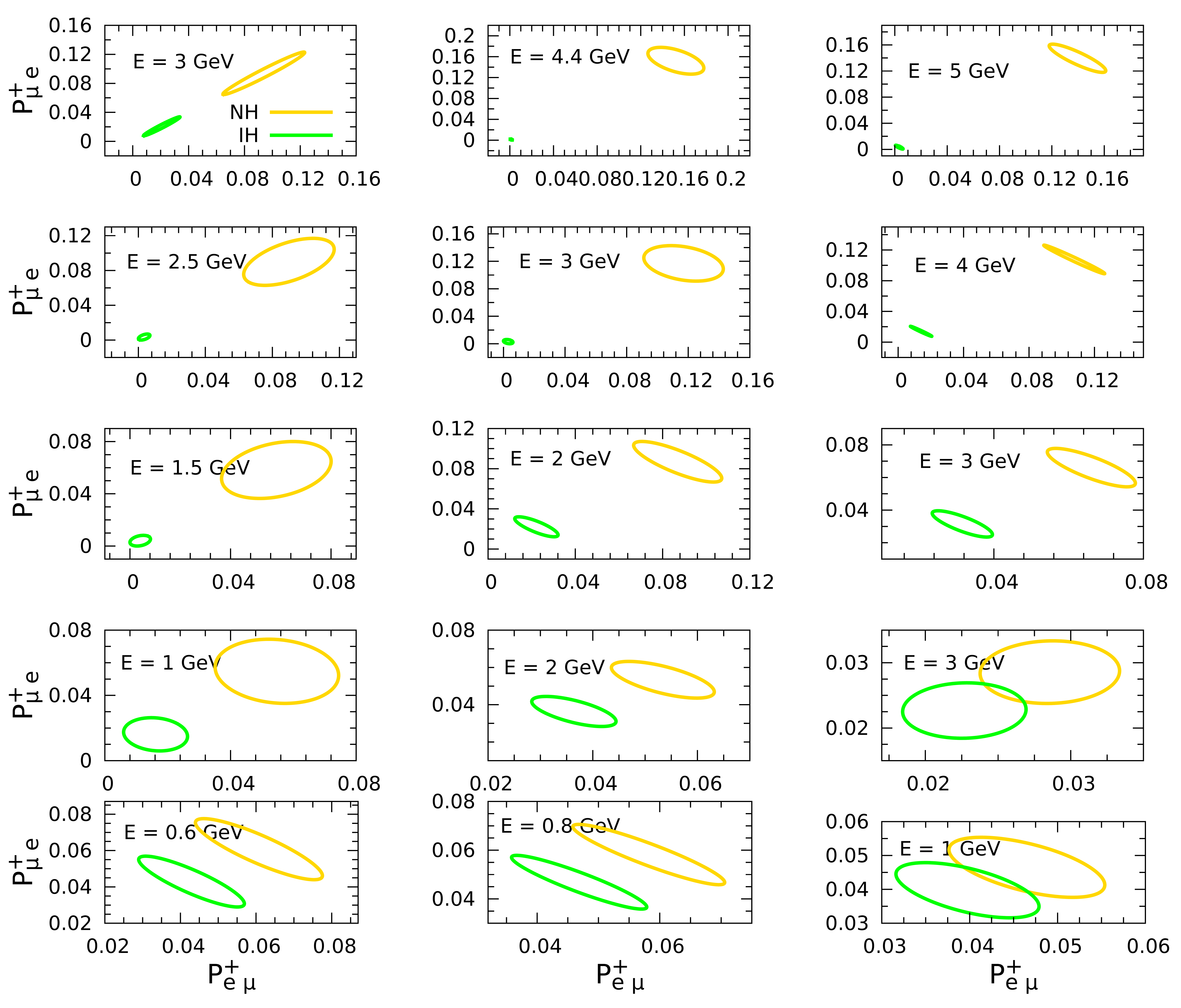}
\caption{This figure illustrates hierarchy ellipses in biprobability plane for experimental setups viz. 
    Brookhaven -- Cornell (L=350 Km) in fifth row of figures, CERN -- Gran Sasso ( L=730 Km) in fourth row, JHF -- Seoul (L=1200 Km) in
    third row, JHF -- Beijing (L=2100 Km) in second row and  Fermilab -- SLAC (L=2900 Km) in first row of figures. 
    In the constant matter density approximation we have chosen $\rho=3.5 ~gm/cm^3$. All the mixing parameters have been taken at their
    best fit values as tabulated in table \ref{tab:parabf}.}
\label{fig:ellips1}
\end{center}
\end{figure*}

Analysis of Figure~\ref{fig:elpsep} reveals following informations:

At L=2900 Km, below $\approx 3$ GeV (say excluded region) two ellipses may overlap, as is also evident from figure \ref{fig:ellips1}.
At this base line two ellipses can be well separated in the appropriate beam energy (3-10) GeV range 
corresponding to (3-16) units of separation between ellipses along the diagonal, 
hence is of paramount importance for mass hierarchy investigation.

At L=2100 Km two ellipses may overlap below $\approx$2 GeV. Thus here the most appropriate beam energy 
range is (2-10) GeV corresponding to ellipses separation of (2-12) units. Hence here ellipses 
will be more closer than the previous experiment at chosen E value.
 
At 1,200 Km overlapping between ellipses may be expected below $\approx 1.3$ GeV. We expect 
separation range of (1-7) units for (1.3-10) GeV.

At 730 Km we expect overlapping between NH and IH ellipses at E$\le$ 0.8 GeV. At this base line in the appropriate 
beam energy range there is expected separation of (1-5) units.

At 350 Km there is overlapping between the ellipses over whole beam energy range under
consideration except for (0.5-1.0) GeV, with expected separation of (1-3) units.

We can conclude from figures \ref{fig:ellsep2d} and \ref{fig:elpsep}, at L=2900 and 2100 Km two hierarchy ellipses are well separated, so that 
we expect to have clean signal for the hierarchy investigation. At L=1200 Km though there is detectable 
separation but for base line L=730 Km separation becomes small and at L=350 Km this separation lowers to 
more smaller value, which makes it difficult to investigate mass hierarchy with presently available 
spreads in beam energy sources. \\

\setlength{\arrayrulewidth}{0.31 mm}  
\setlength{\tabcolsep}{2pt}
\renewcommand{\arraystretch}{1.1}
\begin{table}[h]
\centering
\begin{tabular}{p{2.5cm} p{1.7cm} p{1.7cm} p{1.7cm} p{3.5cm}} \hline
 ~~~Scenario &  E &  L & $A_{e \mu}^{h}$  & $A_{e \mu}^{h}$ (E $\pm$ 0.5 GeV)  \\ 
~~~ (n) & (GeV) & (Km) & (units) & (units) \\  \hline
~~~~~1 & 1 & 507  & 2- 3    & 1 - 3   \\
~~~~~1 & 2 & 1014 & 4 - 5   & 3 - 6   \\
~~~~~1 & 3 & 1521 & 6 - 7   & 5 - 8   \\
~~~~~1 & 4 & 2028 & 8 - 9   & 7 - 10  \\  
~~~~~1 & 5 & 2535 & 11 - 12 & 10 - 13 \\ 
~~~~~2 & 1 & 1507 & 2 - 3   & -6 - 1  \\ 
~~~~~2 & 2 & 3014 &   --    &    --   \\ 
\hline
\end{tabular}
\captionsetup{justification=raggedright,
singlelinecheck=false  }
\caption{Separation between nearest edges of NH and IH ellipses (i.e. $A_{e \mu}^{h}$ given in Eqn.~(\ref{nprhr}))
for n=1 and n=2 scenarios at possible $L$ and  $E$ values given by Eqn.~(\ref{condta}).
Last column shows the possible separation between ellipses 
for beam spread of 1 GeV, at given beam energy E.}
\label{tab:tabsep}
\end{table}
\begin{figure*} 
\begin{center}
\caption{This Figure illustrates an ellipse in $P_{e \upmu} - P_{\upmu e}$ plane for experimental 
    configuration L=2900 Km and E=1.7 GeV. All the other mixing angles and mass square differences have been chosen 
    at their best fit values as has been tabulated in table \ref{tab:parabf} and matter density $\rho=3.5 ~gm/cm^3$.}
\includegraphics [width=.65\textwidth] {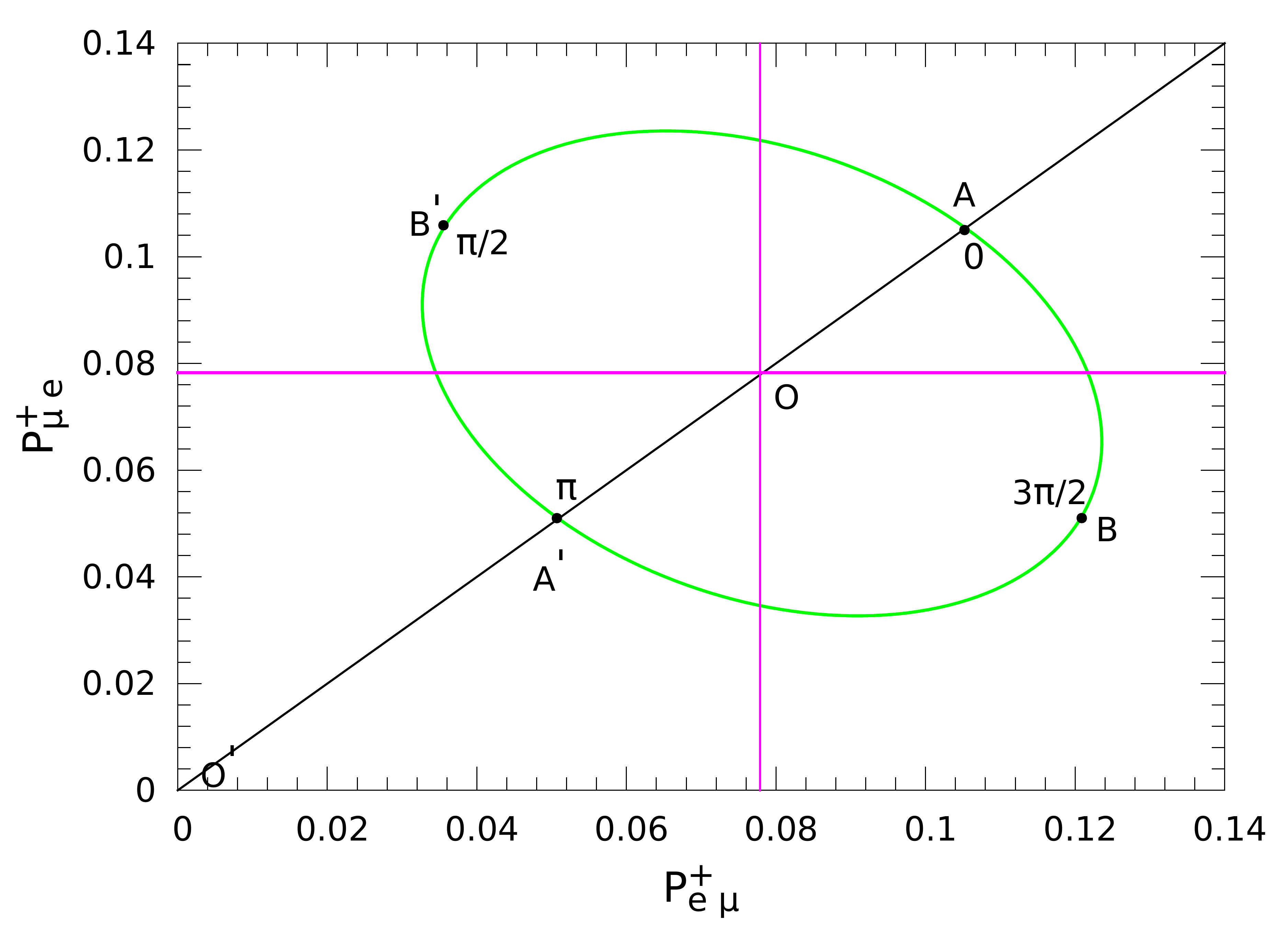} 
\label{fig:ellips2}
\end{center}
\end{figure*}
In Figure~\ref{fig:ellips2} we have depicted an elliptic CP trajectory corresponding to the $0-2~\pi$ variation in 
$\delta^{CP}$ phase at given base line `L' and beam energy `E', drawn in $P_{e \mu}^{+} - P_{\mu e}^{+}$ plane.
We can find with the help of Eqn's.~(\ref{parti1}) and (\ref{partirev}), 
the coordinates of the required points on ellipse as $O(\eta_{1}^{+}~, ~\eta_{1}^{+})$,
$A(\eta_{1}^{+} + \eta_{2}^{+}~, ~\eta_{1}^{+} + \eta_{2}^{+})$ 
and $B(\eta_{1}^{+} + \eta_{3}^{+}~, ~\eta_{1}^{+} - \eta_{3}^{+})$. Now it is easy to find the widths of 
major and minor axis of ellipse as
\begin{subequations}
\begin{eqnarray}
 OA &=& \sqrt{2} ~\eta_{2}^{+} ~~~ so ~that ~~~ A^{'}A = 2 \sqrt{2} ~\eta_{2}^{+} 
 \label{apadis}    \\
 OB &=& \sqrt{2} ~\eta_{3}^{+} ~~~ so ~that ~~~ B^{'}B = 2 \sqrt{2} ~\eta_{3}^{+}
 \label{bpbdis}    \\
 O^{'}O &=& \sqrt{2} ~\eta_{1}^{+}
 \label{opodis}
\end{eqnarray}
\end{subequations} 

\begin{figure*} 
\begin{center}
\caption{ This figure illustrates the parameter $A^{'}A$ given in Eqn.~(\ref{apadis}) in first
     row of figures and parameter $B^{'}B$ given in Eqn.~(\ref{bpbdis}) 
     in the second row, in the E -- L plane. Figures on the left hand side column correspond to 
     normal mass hierarchy (NH) and that on the right hand side to the inverted mass hierarchy (IH). 
     All the other parameters are same as in above figures.}
\includegraphics [width=0.8\textwidth] {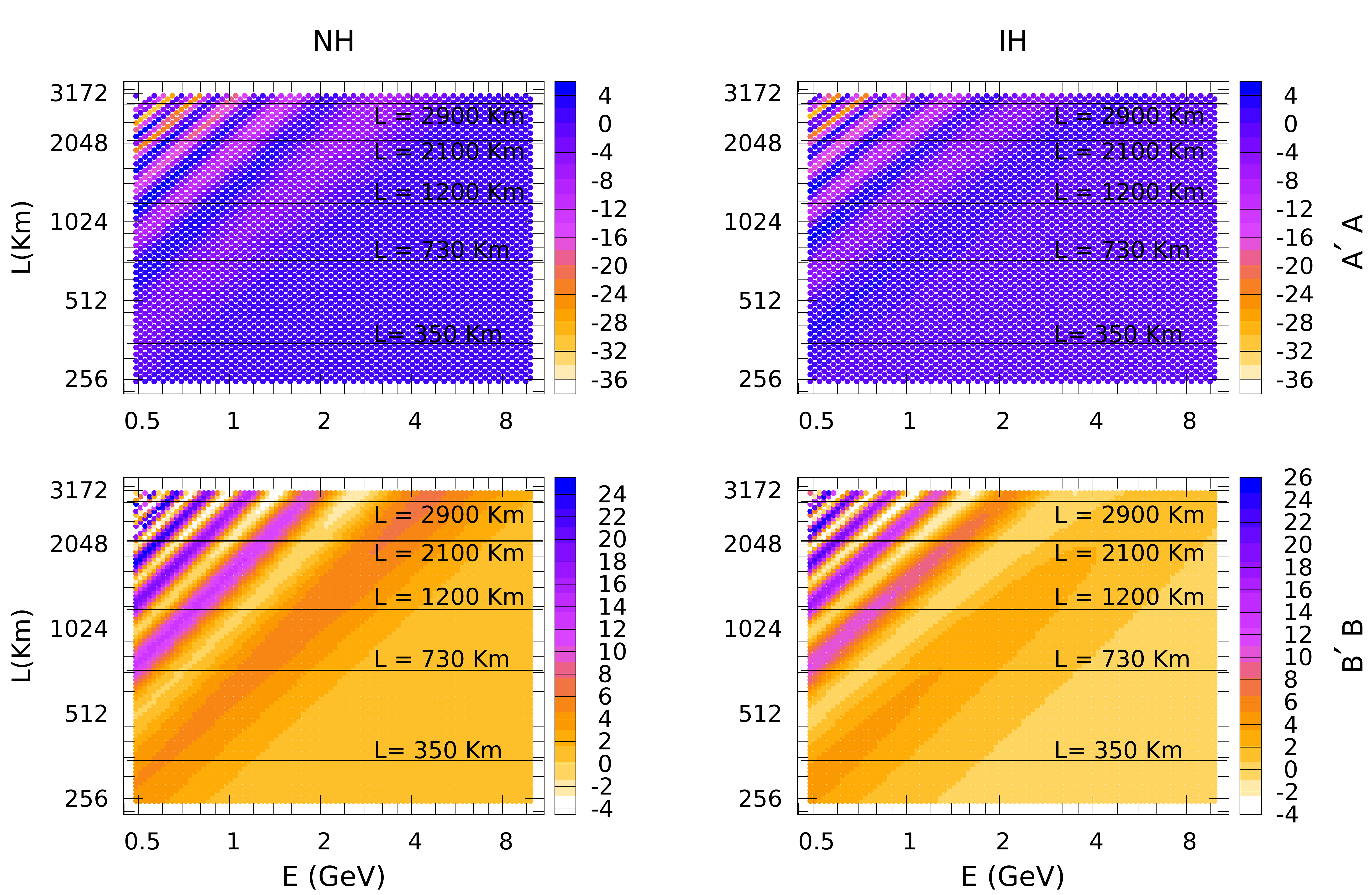} 
\label{fig:majmin}
\end{center}
\end{figure*}

Where $AA^{'}$ for $AA^{'} > BB^{'}$ is the major and for $AA^{'} < BB^{'}$ is the minor axis of ellipse and vice-versa. These two 
axis of ellipse have been drawn in Figure~\ref{fig:majmin} as the contour plots in E-L plane.  

In Figure~\ref{fig:majmin}, we expect to investigate CP-violating phase 
$\delta^{CP}$ narrow ranges with large sized ellipses in the energy region where 
two hierarchy ellipses may overlap in figure \ref{fig:elpsep}, mentioned as excluded energy region in 
the  analysis of Figure~\ref{fig:elpsep} above. But due to fast 
oscillations in this excluded energy ranges and large 
spreads in the currently available beam energy sources,
it is difficult to carry out any experimental 
investigation in these regions presently.
Thus if Figure~\ref{fig:elpsep} helps to determine the optimal experimental configuration while we 
want to investigate mass hierarchy, then Figure~\ref{fig:majmin} is very useful in deciding an optimal 
experiment, when we need large size ellipses in order to investigate CP/T-violation phase ${\delta^{CP}}/{\delta^{T}}$.
Hence if we become sure of the mass hierarchy with the help of Figure~\ref{fig:elpsep},
one can make use of Figure~\ref{fig:majmin} to fix at least narrow range for phase $\delta^{CP}$.

We can conclude to say from above discussion, T-violation bi-probability plots are very promising especially to
the investigation of mass hierarchy.
\section{Parameter degeneracy in $(\theta_{13}, ~\delta^{CP})$ parameter space}
\label{sec:paradgen}
Degeneracy in the parameter space, could create a problem to accurately investigate the mass hierarchy and T-violation phase $\delta^{CP}$,
discussed in above section. If we don't discuss about the parameter degeneracy, our whole discussion will remain incomplete. In 
the reactor mixing angle $\theta_{13}$ and $\delta^{CP}$ parameters space, it is possible that if ($\theta_{13}, ~\delta^{CP}$) is the 
solution of equations (\ref{parti}) and (\ref{partirev}), then ($\theta_{13}^{'}, ~{\delta^{CP}}^{'}$) 
may also be an another solution simultaneously. This is known as problem of parameter degeneracy.
We will further find the possible simultaneous solutions for both the hierarchies and will discuss how to 
get rid of degeneracy problem. A similar type of problem has been already discussed in \cite{min3} and references therein.

In the limit when $\theta_{13}\ll 1$ so that $sin~\theta_{13}\approx \theta_{13} \equiv \theta$ (say), we can rewrite Eqn.~(\ref{parti})
to the form as 
\begin{subequations}
\begin{eqnarray}
 P_{e \mu}^{+} &\equiv& P_{e \mu} \equiv P ~(say)  \nonumber \\
   &=& X^{\pm} ~\theta^{2} ~ + ~ Y^{\pm} ~ cos{(\Delta ~ \mp ~ \delta_{CP})}~\theta ~ + ~ Z^{\pm}
 \label{prb1}
 \end{eqnarray}
 
so that 
\begin{eqnarray}
 P_{\mu e}^{+} &\equiv& P_{\mu e} \equiv P^{T} ~(say)    \nonumber \\
    &=& X^{\pm} ~\theta^{2} ~ + ~ Y^{\pm} ~ cos{(\Delta 
 \pm \delta_{CP})}~\theta ~ + Z^{\pm}
 \label{prb2}
 \end{eqnarray}
 \end{subequations}
 
Here $\Delta_{L}= \Delta \frac{L}{2}$ and coefficients read as
\begin{eqnarray*}
     X^{\pm} = 4~s^{2}_{23} ~\frac{sin^{2}{[(A \mp 1)\Delta_{L}]}}{(A \mp 1)^{2}}      
     \end{eqnarray*}
\begin{eqnarray}
     Y^{\pm} &=& \pm 2 ~\alpha~sin {~2 \theta_{12}} ~sin {~ 2\theta_{23}}~ \frac{sin~{[A \Delta_{L}]}}{A} 
     ~ \frac{sin [(A \mp 1) \Delta_{L}]}{(A \mp 1)}   \nonumber \\                                                     
     Z^{\pm} &\equiv& Z = \alpha^{2} ~sin^{2} {2\theta_{12}}~ c^{2}_{23}~ \frac{sin^{2}{[A \Delta_{L}]}}{A^{2}}
     \label{coefnt}
\end{eqnarray}

where upper ` + ' sign corresponds to +ve sign of $\Delta m_{31}^{2}$ i.e. case of Normal mass 
Hierarchy (NH) and lower ` - ' sign to -ve sign of $\Delta m_{31}^{2}$ i.e. case of Inverted mass Hierarchy (IH).

Now for any given point $(P, ~P^{T})$ in the $P_{e \mu} -  P_{\mu e}$ plane, one can write solution to 
Eqn's.~(\ref{prb1}) and (\ref{prb2}) as 
\begin{eqnarray}
 \theta &=& (\pm) \sqrt{\frac{P ~ - ~Z^{\pm}}{X^{\pm}}} ~ - ~\frac{Y^{\pm}}{2 ~X^{\pm}} ~cos(\Delta \mp \delta_{CP}) 
    \nonumber \\
and &&   \nonumber \\
 \theta^{T} &=& (\pm) \sqrt{\frac{P^{T} ~ - ~Z^{\pm}}{X^{\pm}}} - \frac{Y^{\pm}}{2 ~X^{\pm}} ~cos(\Delta \pm \delta_{CP})
 \label{soln}
\end{eqnarray}

It is possible that if an ellipse $(\theta_{1},~\delta^{CP}_{1})$ corresponds to point $(P, ~P^{T})$, then an 
another ellipse say $(\theta_{2},~\delta^{CP}_{2})$ may also correspond to same point. Note, $\delta^{CP}_{i}$ denotes 
different values of CP--violating phase $\delta_{CP}$. Thus we can write difference as
\begin{eqnarray}
 \theta_{2} - \theta_{1} &=& - ~\frac{Y^{\pm}}{2 ~X^{\pm}} \left[ \left( cos~\delta^{CP}_{2} - cos~\delta^{CP}_{1} \right)
 ~cos~\Delta \right. \nonumber \\
 && \left. \pm ~\left( sin~\delta^{CP}_{2} - sin~\delta^{CP}_{1} \right) ~sin~\Delta \right]
 \nonumber  \\
and ~~~ \theta^{T}_{2} - \theta^{T}_{1} &=& - ~\frac{Y^{\pm}}{2 ~X^{\pm}} \left[ \left( cos~\delta^{CP}_{2} - cos~\delta^{CP}_{1} \right)
 ~cos~\Delta  \right.  \nonumber \\ 
   &&  \left.  \mp ~ \left( sin~\delta^{CP}_{2} - sin~\delta^{CP}_{1} \right) ~sin~\Delta \right]
 \label{soln1}
\end{eqnarray}

In this equation we have $\theta^{T}_{i} = \theta_{i}$ with {\it i}=1,2 as both belong to same ellipse in 
$P_{e \mu} -   P_{\mu e}$ plane.  
Which entails the degeneracy that if $(\theta_{1}, \delta^{CP}_{1})$ is the solution so is the 
\begin{eqnarray} 
 \theta_{2}= \theta_{1} ~ + ~\frac{Y^{\pm}}{X^{\pm}} ~cos~\delta^{CP}_{1} ~cos~\Delta ~~ and ~~ 
  \delta^{CP}_{2}=\pi - \delta^{CP}_{1}
  \label{soln3}
\end{eqnarray}

Thus in total we have four degenerate solutions as 
\begin{eqnarray}
 (\theta_{1}, ~\delta_{1}^{CP})[NH] &\equiv& (\theta_{1}, ~\delta_{1}^{CP})  \nonumber  \\
 (\theta_{2}, ~\delta_{2}^{CP})[NH] &\equiv& (\theta_{2}, ~\delta_{2}^{CP}) ~~~ with ~~ \delta_{2}^{CP} = \pi - \delta_{1}^{CP} 
 \nonumber  \\
 (\theta_{1}, ~\delta_{1}^{CP})[IH] &\equiv& (\theta_{3}, ~\delta_{3}^{CP})  \nonumber  \\ 
 (\theta_{2}, ~\delta_{2}^{CP})[IH] &\equiv& (\theta_{4}, ~\delta_{4}^{CP}) ~~~ with ~~ \delta_{4}^{CP} = \pi - \delta_{3}^{CP}
 \label{soln4}
\end{eqnarray}

It is clear that same sign solutions are well connected i.e. if we know one solution, we can find the other, but not the opposite sign ones. 
To find relation between opposite sign solutions we can have from Eqn.~(\ref{soln})
\begin{eqnarray}
\theta_{1} &=& (\pm) \sqrt{\frac{P ~ - ~Z}{X^{+}}} ~ - ~\frac{Y^{+}}{2 ~X^{+}} ~cos\left( \Delta - \delta^{CP}_{1} \right) 
 \nonumber  \\
 \theta_{1}^{T} &=& (\pm) \sqrt{\frac{P^{T} ~ - ~Z}{X^{+}}} ~ - ~\frac{Y^{+}}{2 ~X^{+}} ~cos\left( \Delta + \delta^{CP}_{1}\right) 
 \label{soln5}  \\
 \theta_{3} &=& (\pm) \sqrt{\frac{P ~ - ~Z}{X^{-}}} ~ - ~\frac{Y^{-}}{2 ~X^{-}} ~cos\left( \Delta + \delta^{CP}_{3}\right) 
 \nonumber  \\
 \theta_{3}^{T} &=& (\pm) \sqrt{\frac{P^{T} ~ - ~Z}{X^{-}}} ~ - ~\frac{Y^{-}}{2 ~X^{-}} ~cos\left( \Delta - \delta^{CP}_{3}\right)
 \label{soln6}
\end{eqnarray}

In above equation we have chosen $Z^{+} = Z^{-} \equiv Z$  (say), as is evident from Eqn.~(\ref{coefnt}).

Now by $\theta_{1}-\theta_{1}^{T}=0$ gives us following relation 
\begin{eqnarray}
 sin ~\Delta ~sin ~\delta^{CP}_{1} = (\pm) \frac{\sqrt{X^{+}}}{Y^{+}} \left( \sqrt{P ~ - ~Z} ~ - ~\sqrt{P^{T} ~ - ~Z} \right)
 \label{soln7}
\end{eqnarray}

and $\theta_{3}-\theta_{3}^{T}=0$ gives the following relation
\begin{eqnarray}
 sin ~\Delta ~sin ~\delta^{CP}_{3} = (\mp) \frac{\sqrt{X^{-}}}{Y^{-}} \left( \sqrt{P ~ - ~Z} ~ - ~\sqrt{P^{T} ~ - ~Z} \right)
 \label{soln8}
\end{eqnarray}

Now by making use of identity \cite{min3},
\begin{eqnarray}
 \frac{\sqrt{X^{+}}}{Y^{+}} = - ~\frac{\sqrt{X^{-}}}{Y^{-}}
 \label{identy}
\end{eqnarray}

We can find by making use of equations (\ref{soln7}) \& (\ref{soln8}) the following relation,
\begin{eqnarray}
 sin ~\delta_{1}^{CP} = sin ~\delta_{3}^{CP}
 \label{opstsn}
\end{eqnarray}

Thus one can choose without loss of generality $\delta_{3}^{CP} = \uppi - \delta_{1}^{CP}$. Now 
by making use of this value of $\delta_{3}^{CP}$, we can write from Eqn's.~(\ref{soln5}) and (\ref{soln6}),  
\begin{eqnarray}
 \theta_{1} &=& (\pm) \frac{\left( \sqrt{P ~ - ~Z} ~ + ~\sqrt{P^{T} ~ - ~Z} \right)}{2 ~\sqrt{X^{+}}} 
 ~ - ~\frac{Y^{+}}{2 ~X^{+}} ~cos ~\Delta ~cos ~\delta^{CP}_{1}  
 \nonumber  \\
 \theta_{3} &=& (\pm) \frac{\left( \sqrt{P ~ - ~Z} ~ + ~\sqrt{P^{T} ~ - ~Z} \right)}{2 ~\sqrt{X^{-}}} 
~ - ~\frac{Y^{-}}{2 ~X^{-}} ~cos ~\Delta ~cos ~\delta^{CP}_{3}  \nonumber \\
 \label{soln9}
\end{eqnarray}
Above equation can be solved further by making use of Eqn.~(\ref{identy}) to following form,
\begin{eqnarray}
 \sqrt{X^{+}} ~\theta_{1} ~ - ~\sqrt{X^{-}}~\theta_{3} = - ~\frac{Y^{+}}{2 \sqrt{X^{+}}} \left( cos ~\delta^{CP}_{1}  
 + cos ~\delta^{CP}_{3} \right) ~cos ~\Delta  \nonumber \\
 \label{rel13}
\end{eqnarray}
We can further solve above equation with the help of Eqn.~(\ref{opstsn}) to get,
\begin{eqnarray}
 \theta_{3} = \sqrt{\frac{X^{+}}{X^{-}}} ~ \theta_{1}  ~~~ and ~~~ \delta_{3}^{CP} = \pi - \delta_{1}^{CP}.
 \label{thta3}
\end{eqnarray}
Now we can write from Eqn's.~(\ref{soln3}), (\ref{soln4}) and (\ref{thta3}) the four degenerate solutions 
corresponding to equations (\ref{prb1}) and (\ref{prb2}) explicitly as  
\begin{eqnarray}
 For  &&  (\theta_{1}, ~\delta_{1}^{CP})   \hspace{0.5cm} chosen  \nonumber \\
 \theta_{2} &=& \theta_{1} ~ + ~\frac{Y^{+}}{X^{+}} ~cos~\Delta ~cos~\delta^{CP}_{1}  \hspace{0.2cm} and \hspace{0.2cm} 
  \delta^{CP}_{2}=\pi - \delta^{CP}_{1} \nonumber  \\
 \theta_{3} &=& \sqrt{\frac{X^{+}}{X^{-}}} ~ \theta_{1}  \hspace{2.3cm} and \hspace{0.2cm} \delta_{3}^{CP} = \pi - \delta_{1}^{CP}  
 \nonumber  \\
   \theta_{4} &=& \theta_{3} ~ - ~\frac{Y^{-}}{X^{-}} ~cos~\Delta ~cos~\delta^{CP}_{1}  \hspace{0.2cm} and \hspace{0.2cm} 
   \delta^{CP}_{4} =\pi - \delta^{CP}_{3} = \delta_{1}^{CP} \nonumber \\
  \label{soln11}
\end{eqnarray}
\begin{figure*} 
\caption{\label{fig:ellips3} In this figure we have chosen $(P_{e \mu}^{+}, ~P_{\mu e}^{+})= (4.3, ~5.6) \%$ in case of LHS Fig. (i.e. for L=350 Km, E=1 GeV) and 
    $(P_{e \mu}^{+}, ~P_{\mu e}^{+})= (3.5, ~5.6) \%$ in case of RHS Fig. (i.e. for L=730 Km, E=1 GeV) shown by red dot. 
    This red dot represents the four degenerate solutions corresponding 
    to equations (\ref{prb1}) and (\ref{prb2}).} 
\includegraphics [width=0.99\textwidth] {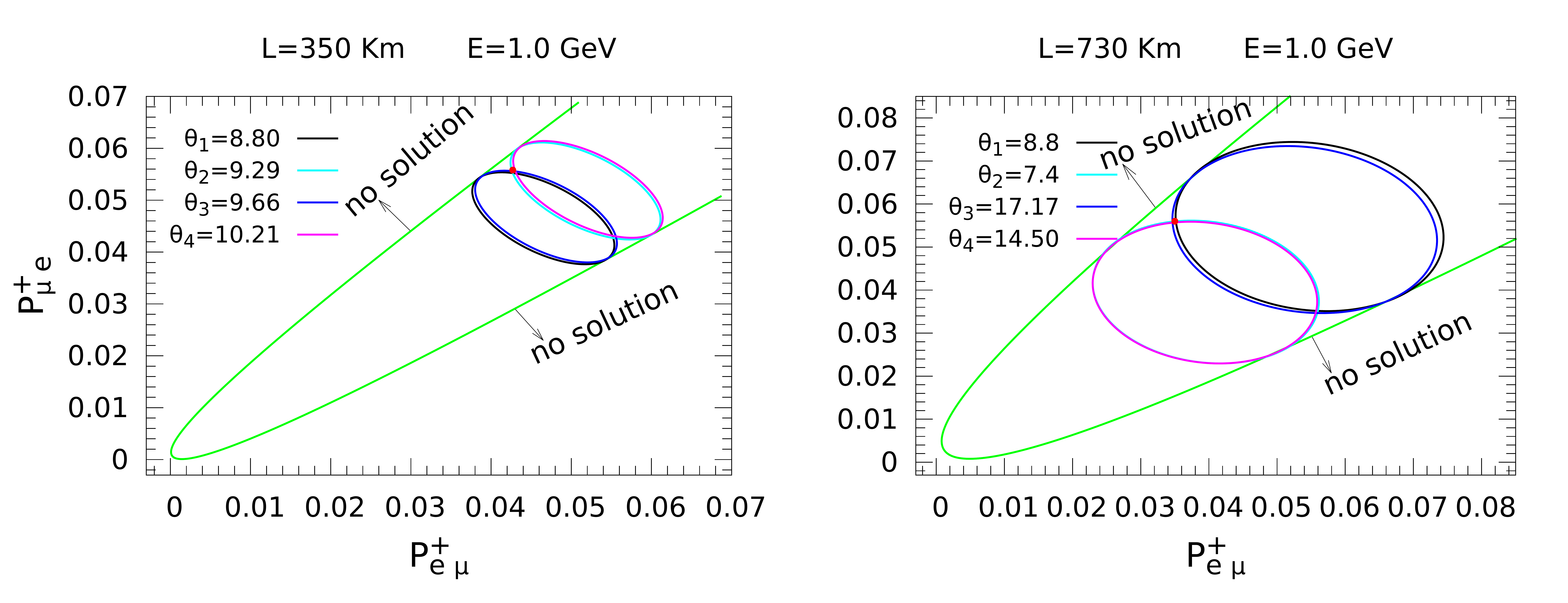}
\end{figure*}
In the $P_{e \mu}^{+} - P_{\mu e}^{+}$ plane, physically allowed region is restricted by the condition $sin^{2} \delta^{CP}_{i}\leq 1$, 
from which we can formulate with the help of Eqn's.~(\ref{soln7}) and (\ref{soln8}) the following inequality,
\begin{eqnarray}
\left( \sqrt{P ~ - ~Z} ~ - ~ \sqrt{P^{T} ~ - ~Z} \right)^{2} \leq \frac{{Y^{+}}^{2}}{X^{+}} ~sin^{2}\Delta
= \frac{{Y^{-}}^{2}}{X^{-}} ~sin^{2}\Delta 
\label{phsrgn}
\end{eqnarray}
This physical region has been shown by green curve enclosing the ellipses in Figure~\ref{fig:ellips3}.
In this Figure we note that at small base line, L=350 Km (Figure on LHS) four solutions lie in the 3 $\sigma$
range $(7.7 \le \theta_{13}^{o} \le 9.9)$ \cite{parabf} except the $\theta_{4}= 10.21^{o}$.
Similarly at L=730 Km (Figure on RHS), two solutions $\theta_{3}= 17.17^{o}$
and $\theta_{4}= 14.5^{o}$ lie outside the 3 $\sigma$ range of third mixing 
angle, which suggests that inverted mass hierarchy signal is well distinguishable even at 3 $\sigma$ level for this 
experimental configuration. Thus we conclude to say, if there is opposite sign degeneracy at small base line,
it will get circumvented at longer base lines for the same other experimental inputs. This is due to the fact, as matter effects
in long base lines increase, which in turn pull the two different hierarchy CP trajectory ellipses more apart,
as we can confirm from Figure~\ref{fig:ellips1}. 
Opposite sign degeneracy has physical manifestation as long as two ellipses
corresponding to NH and IH are overlapping. We can avoid overlapping of 
opposite sign ellipses by appropriately choosing longer base line L and specific beam energy E.
To find an optimized L and E configuration, Figure~\ref{fig:elpsep} is very guiding.

The same sign degeneracy can be removed by choosing L and E such that 
\begin{subequations}
\begin{eqnarray}
cos ~\Delta_{L} = 0, ~~~~ i.e. ~\Delta_{L}=(2n-1) \uppi/2.
\label{samdgn}
\end{eqnarray}
This condition is fulfilled for 
\begin{eqnarray}
L &\simeq& 507 \frac{E}{(GeV)} ~Km ~for ~n=1  \nonumber  \\ 
and ~~ at ~L &\simeq& 1507 ~\frac{E}{(GeV)} ~Km ~for ~n=2.
\label{condta}
\end{eqnarray}
\end{subequations}

Possible configurations of experiments for n=1 and n=2 scenarios have been given in Table~\ref{tab:tabsep}
in Section~\ref{sec:phenology} above.  

\section{Conclusions and perspectives}

To explore an optimal set up in order to investigate mass hierarchy, separation between two 
opposite hierarchy ellipses in the biprobability (P -- $P^{T}$) plane,
serves as an optimization parameter. It is evident from 
Figures~\ref{fig:elpsep} and \ref{fig:ellips1}, at larger base lines 
hierarchy ellipses separate to large extent as compared to the smaller base lines, which suggests that 
longer base lines are useful to investigate mass hierarchy in the the $P-P^{T}$ plane. We come to know from 
figures \ref{fig:ellsep2d} and \ref{fig:elpsep}
that at given base line below a specific value of beam energy, oscillations becomes very fast and separation between 
hierarchy ellipses becomes so small, that they may overlap to large extent too. This value of beam energy has values $\lesssim$ 
3, 2, 1.3, 0.8 and $\lesssim$ 0.1 \& $\gtrsim$ 1.0 GeV, respectively for base lines 2900, 2100, 1200, 730 and 350 Km.

The hierarchy investigation in the context of biprobability-plane provides an elegant way. The major 
drawback that could hinder precise measurement of hierarchy is that,
there always exist four fold degeneracy for any coordinate in the $P-P^{T}$ plane within the $(\theta_{13}, ~\delta_{CP})$
parameter space. Opposite sign degeneracy have physical existence in the case when two hierarchy ellipses overlap. 
To get rid of this degeneracy we need to observe the results at suitable long base lines and beam energy.
Same sign degeneracy disappears as soon as $cos ~\Delta_{L} = 0$, which is fulfilled for $L \simeq 507 ~\frac{E}{(GeV)}$ Km for n=1 
and for $L \simeq 1507 \frac{E}{(GeV)}$ Km for n=2 scenario. It is recommended to especially follow Figure~\ref{fig:elpsep} in order to 
choose suitable L and E configurations at which both the opposite and same sign degeneracies can be circumvented.
Also we can analyze from Table~\ref{tab:tabsep} that, experimental configuration L=2,535 Km, E=5 GeV for 
n=1 scenario, is the most suitable one that fulfills the criteria mentioned in previous line.




\begin{thebibliography}{10}
%
\bibitem{t2k}  R.  Patterson   (NO$\nu$A     Collaboration), Nucl. Phys. Proc. Suppl. 235-236, 151 (2013);
K. Abe et al. (T2K Collaboration), Nucl. Instrum. Meth. A {\bf 659,} 106 (2011); K. Abe et al.,  
Phys. Rev. Lett. {\bf 107,} 041801 (2011). 


\bibitem{lbne} C. Adams et al. (LBNE Collaboration), arXiv:1307.7335 [hep-ex]; A. Stahl et al.,CERN-SPSC-2012-021, SPSC-EOI-007.
     
\bibitem{jhf} 
Y. Itow et al., ``The JHF-Kamioka neutrino project,'' \textbf{hep-ex/0106019}.

\bibitem{minos}
MINOS Collaboration, P. Adamson, et al., MINOS Detectors Technical Design Report, Version 1.0, NuMI-L-337, 1998.

\bibitem{opera}
OPERA Collaboration, M. Guler, et al., ``OPERA: An Appearance
Experiment to Search for Nu/Mu $\leftrightarrow$ Nu/Tau Oscillations
in the CNGS Beam. Experimental Proposal,'' CERNSPSC-
2000-028, CERN-SPSC-P-318, LNGS-P25-00, 2000.     

\bibitem{seri}
Evgeny K. Akhmedov, Robert Johansson, Manfred Lindner,
Tommy Ohlsson, and Thomas Schwetz, JHEP \textbf{04} (2004) 078, \textbf{hep-ph/0402175}.



\bibitem{wolf}
L. Wolfenstein, Phys. Rev. D {\bf 17}, 2369 (1978); ibid. 20, 2634 (1979).

\bibitem{simrn}
S.P. Mikheyev and A.Yu. Smirnov, Yad. Fiz. {\bf 42}, 1441 (1985) [Sov. J. Nucl. Phys. {\bf 42},913 (1985)]; I1 Nuovo Cim.
C 9, 17 (1986); Zh. Eksp. Teor. Fiz. {\bf 91}, 7 (1986) [Sov. Phys. JETP \textbf{64}, 4 (1986)].

\bibitem{lit1}
K. Kimura, A. Takamura, H. Yokomakura, Physics Letters B {\bf 537}, 1 (2002), \textbf{hep-ph/0203099}; Keiichi Kimura, Akira Takamura, 
     Hidekazu Yokomakura, Phys. Rev. D {\bf 66}, 073005 (2002); Hidekazu Yokomakura, Keiichi Kimura, Akira Takamura,
     Physics Letters B {\bf 544}, 3 (2002), \textbf{ hep-ph/0207174}; A. Cervera, A. Donini, M.B. Gavela, J.J. Gomez C\'{a}denas,
     P. Hern\'{a}ndez, O. Mena and S. Rigolin, Nucl.Phys. B {\bf 579} (2000) 17, \textbf{hep-ph/0002108}.
     
\bibitem{lit2}
Tommy Ohlsson, Physica Scripta. Vol. {\bf T93}, 18, 2001; Tommy Ohlsson and H$\mathring{a}$kan Snellman, 
    J.Math.Phys. 41 (2000) 5, J.Math.Phys. 42 (2001) 2345, \textbf{hep-ph/9910546}.     

\bibitem{lit3}
S. M. Bilenky, C. Giunti and W. Grimus, Progress in Particle and Nuclear Physics 43, 1 (1999).

\bibitem{lit4}
Jiro Arafune and Joe Sato, Phys. Rev .D {\bf55}, 1653 (1997); Jiro Arafune, Masafumi Koike, and Joe Sato, Phys. Rev. D {\bf 56}, 3093 (1997).

\bibitem{lit5}
C. Jarlskog, Phys. Rev. Lett. 55, 1039 (1985); Z. Phys. C 29, 491 (1985); C. Jarlskog and R. Stora, Phys. Lett.
B {\bf 208}, 268 (1988); C. Jarlskog, Proc. of CP Violation, edited by C. Jarlskog, Advanced Series in High Energy
Physics Vol. 3, p. 3, World Scientific, Singapore, 1989.

\bibitem{lit6}
L.L. Chan and W.Y. Keung, Phys. Rev. Lett. \textbf{53}, 1802 (1984).

\bibitem{lit7}
R.N. Mohapatra and G. Senjanovid, Phys. Rev. D \textbf{23}, 165 (1981).




\bibitem{basym}
M. Fukugita and T. Yanagida, Phys. Lett.B \textbf{ 174}, 45 (1986).

\bibitem{th13n}
P. Huberet al., Phys. Rev. D {\bf 70}, 073014 (2004), \textbf{ hep-ph/0403068}.

\bibitem{tvilekh}
E. K. Akhmedov, P. Huber, M. Lindner, and T. Ohlsson, Nucl. Phys. B \textbf{608}, 394 (2001),
\textbf{hep-ph/0105029}.

\bibitem{tvilpet}
P. I. Krastev and S. T. Petcov, Phys. Lett. B \textbf{205}, 84 (1988).

\bibitem{pdegn1}
H. Minakata and H. Nunokawa, JHEP \textbf{0110} (2001) 001 [hep-ph/0108085]; H. Minakata and H. Nunokawa,
J. High Energy Phys. \textbf{10}, 001~(2001).

\bibitem{pdegn2}
J. Burguet-Castell, M.B. Gavela, J.J. Gomez-Cadenas, P. Hernandez and O. Mena,
Nucl. Phys. B \textbf{ 608} (2001) 301; H. Minakata, H. Nunokawa, and S.J. Parke, Phys. Lett. B \textbf{537},
249 ~(2002).

\bibitem{th13a}
Kwong Lau, Status of $\theta_{13}$ measurement in reactor experiments, \textbf{arXiv:1308.0089}, (2013).

\bibitem{th13b}
L. J. Halla and G. G. Ross, \textbf{arXiv:1308.0089}, (2013); 
F. An et al. (Daya Bay Collaboration), Phys. Rev. Lett. {\bf 108,} 171803 (2012); 
J. Ahn et al. (RENO collaboration), Phys. Rev. Lett.
    {\bf 108,} 191802 (2012).

\bibitem{cptnu}
Hisakazu Minakata, Hiroshi Nunokawab and Stephen Parke, \textbf{hep-ph/0306221} (2003).

\bibitem{min3}
Hisakazu Minakata, Hiroshi Nunokawa and Stephen Parke, Phys. Rev. D \textbf{ 66}, 093012 ~(2002).

\bibitem{parabf}
D. V. Forero, M. T$\acute{o}$rtola and J. W. F. Valle, \textbf{arXiv:1405.7540} (2014).



\end{thebibliography}
\end{document}